\documentclass[preprint]{aastex}
\usepackage{psfig}
\def\Sec{${}^{\prime\prime}$\llap{.}}

\def\teff{T$_{\rm eff}$~}

\begin{document}
\title{Abundances of Extremely Metal-Poor Star Candidates\footnote{
The data presented herein were obtained at the W.M. Keck Observatory,
which is operated as a scientific partnership among the California
Institute of Technology, the University of California and the
National Aeronautics and Space Administration. The Observatory was
made possible by the generous financial support of the W.M. Keck
Foundation.}}

\author{David K. Lai, Michael Bolte, Jennifer A. Johnson{\altaffilmark{2}}, and Sara Lucatello{\altaffilmark{3}}}
\affil{UCO/Lick Observatory, University of California, Santa Cruz,
CA~95064}
\email{david@ucolick.org, bolte@ucolick.org,
Jennifer.Johnson@nrc-cnrc.gc.ca, lucatello@pd.astro.it}
\altaffiltext{2}{DAO/HIA/NRC, 5071 West Saanich RD, Victoria, BC V9E 2E7}
\altaffiltext{3}{INAF-OAPD vicolo dell'Osservatorio 5, 35122 Padova Italy}

\begin{abstract}
We present chemical abundances for 110 stars identified in objective-prism
surveys as candidates to be very metal-poor. The abundances are derived
from high S/N, intermediate-resolution spectra obtained with the Keck
Observatory Echelle Spectrometer and Imager.  An additional 25 stars
with well-determined abundances ranging from [Fe/H]$=-1.5$ and $-3.2$
were observed and the results used to help calibrate our analysis and
determine the accuracy of our abundance determinations. Abundances for
the program stars were measured for Fe, Mg, Ca, Ti, Cr and Ba with an
accuracy of approximately 0.3 dex.  53 of the stars in our sample have
[Fe/H]$\leq -2$, 22 have [Fe/H]$\leq -2.5$ and 13 stars have
[Fe/H]$\leq -2.9$. Surprisingly, approximately one third of the sample is
relatively metal rich with [Fe/H]$>-1.5$.
In addition to identifying a number of extremely metal-poor stars,
this study also shows that moderate-resolution spectra obtained
with the Keck Echelle Spectrometer and Imager yield relatively accurate 
abundances for stars as faint as $V=14$ in modest exposure time ($\sim 20$ minutes).
This capability will prove useful if the so-far elusive stars at [Fe/H]$<-4$
turn out to be mostly fainter than $V=15$.

\end{abstract}


\section{Introduction}\label{intro}
In the last decade the number of identified extremely metal-poor (EMP)
stars with [Fe/H]$<-2.5$ has
increased by almost an order of magnitude. Studies of these relics of
the early star-formation epoch of the Galaxy have been used for
investigations ranging from deriving ages for individual halo stars to
improving our knowledge of neutron capture cross-sections in heavy
nuclei.  The most extreme cases, stars with [Fe/H]$<-3$, should carry the
signatures of only a few post-Big Bang nucleosynthetic events. 
One of
the exciting prospects for investigations in this area is to
identify the nature of ``Population III'' objects through their
nucleosynthetic signatures in the atmospheres of [Fe/H]$<-3$ stars.

The majority of
the known stars with [Fe/H]$<-3$ were discovered as part of the ``HK''
survey of Beers, Preston \& Shectman (1992) (hereafter BPS).
Candidates were identified in objective-prism survey plates
obtained with either the Burrell Schmidt telescope located at Kitt Peak (leading
to the candidates with BS designations) or the Curtis Schmidt telescope located
at Cerro Tololo (leading to the candidates with CS designations).
Follow-up spectra obtained at low
spectral resolution were then used to make rough [Fe/H] estimates and
further refine the sample. The objects identified through this process 
have become the targets for a large number of followup studies
with large telescopes and high-resolution spectrometers starting with
Mcwilliam et al. (1995) and continuing through the present
(e.g.  Cayrel et al., 2004).

Many of the BPS candidates from the Curtis Schmidt survey have had fairly secure metal abundances determined.  In this paper we present the results of analysis of 
110 stars mostly from the BPS survey for metal-poor stars.  Many of the candidates from the BS survey have only photometry published.  In order to determine [Fe/H] with reasonable accuracy, 
we obtained spectra of these stars along with those for a number of well-studied
abundance ``standards'' with the Echelle
Spectrometer and Imager (ESI) at the Keck 2 telescope (Sheinis et al.
2002, and see brief description below). This spectrometer, with relatively short exposures, produces spectra with high enough resolution and signal-to-noise to measure the abundances of a range of elements from Mg to Ba.
 In the following sections we discuss  the effectiveness and
limitations of using ESI for measuring abundances in metal-poor stars and present
equivalent-width-based abundances for the program stars.  Our principal goal is to identify new EMP stars.

\section{Observations} 
Over several runs and four years, we used ESI to obtain spectra of 110
stars from the lists of BPS and Norris, Beers \& Ryan (1999).  We also
observed 25 metal-poor stars for which high spectral resolution, high
quality studies exist in the literature to evaluate the
suitability of ESI spectra
for measuring chemical abundances. We will refer to stars in this 
latter group as
``abundance standards'' for the remainder of the paper.  Twelve of the
standards were chosen from the objects studied in Fulbright (1999), three from Pilachowski et al.
(1996) and three are the globular cluster giants M15-S4, M15-K341, and
M92 V{\small II}-18 with abundance
studies from Kraft \& Ivans (2003), Sneden et al. (1997), and
Sneden et al. (2000).  An additional eight of the BS/CS stars with published high
spectral resolution studies were also used as abundance standards.  

We chose the BS/CS program objects by targeting the most metal-poor stars
from Table 5 of BPS and by selecting the stars from Norris, Beers \&
Ryan (1999) that showed the largest $U-B$ excess at $B-V$ colors
$>0.5$.

We used ESI in the echellete mode with the 0\Sec75-wide slit.  This
results in a spectral resolution of $R \simeq 7000$ and wavelength
coverage from 3900 to 10000 \AA{}. Exposure times ranged from 300 to 1200
seconds.  Star designation from the above references, date of
observation, exposure time and signal-to-noise at 6035 \AA{} are
presented in Table \ref{photometry} for the standards.  Table
\ref{progphot} has the observation details for the program stars, all
of which have signal-to-noise values of around 150 per pixel.

\section{Data Reduction and Extraction}

\subsection{Processing \label{process}}

\begin{figure}
\begin{center}
\scalebox{.75}[.75]{
\plotone{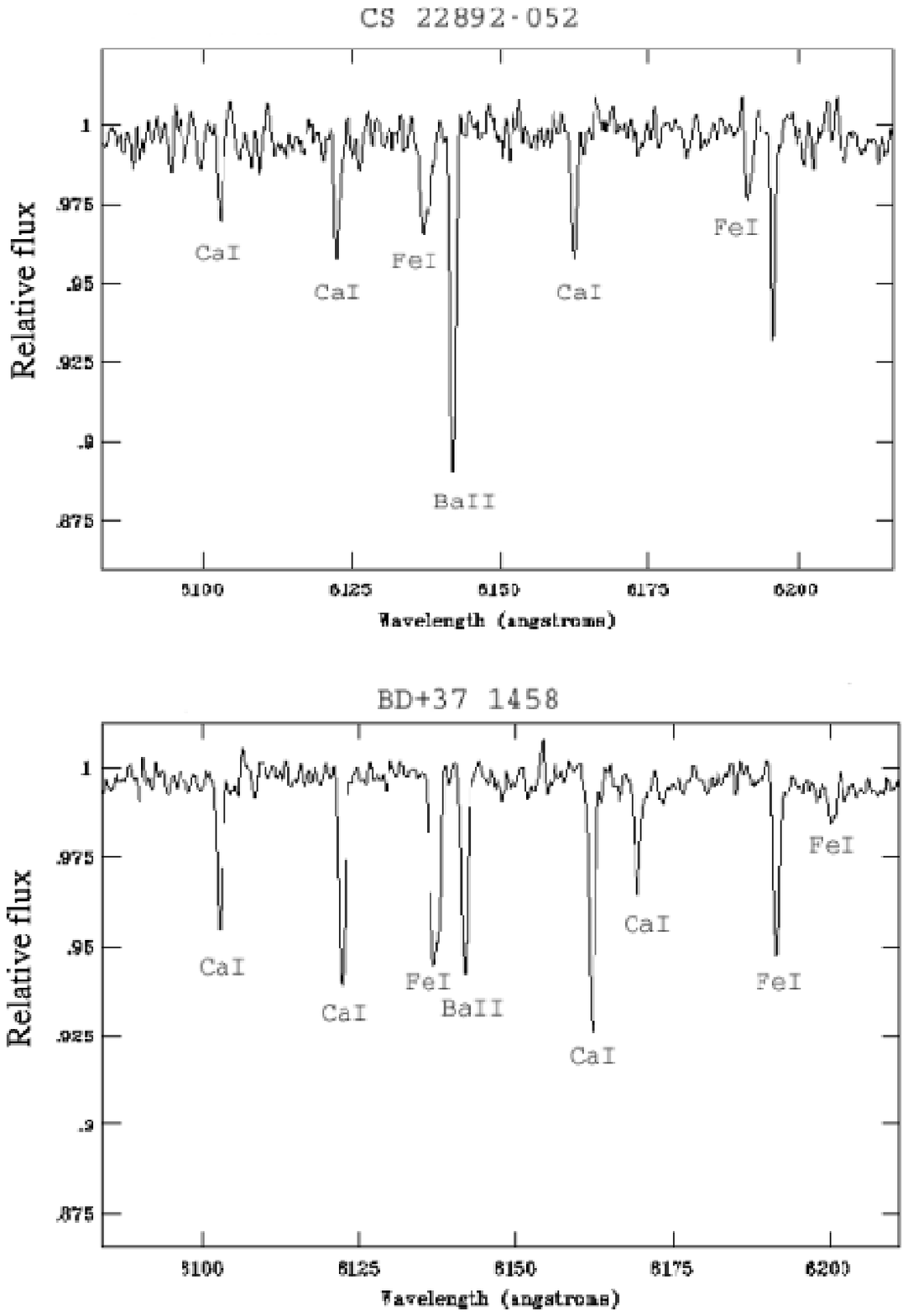}}
\end{center}
\figcaption[Lai.fig1.eps]{Sample regions of two spectra.  Both have had been
corrected to rest frame velocity and identified features have been
labeled. \label{spectra}}
\end{figure}

Image preparation and reduction was carried out using IRAF.  The
spectra were extracted using the echelle package.  Scattered light was
fit and subtracted with a smooth function calculated from the regions
between the spectral orders using the APSCATTER routine.  The routine
APALL was then used to extract the orders into one-dimensional
spectra.  The spectra were then normalized using the CONTINUUM routine
in IRAF.  Although arcs were taken during each run, we found it
adequate to use only a single wavelength solution for all of the
spectra and make a small (generally less than 5 km s$^{-1}$) zeropoint 
correction based on the positions of night sky lines.
A sample region from two spectra
is shown in Figure \ref{spectra}.

\subsection {Equivalent Width Measurements and the Spectral Line List
\label{ewsection}}

The program SPECTRE (Fitzpatrick \& Sneden 1987) was used to measure
equivalent widths of the spectral lines in the extracted, continuum-normalized
spectra.
In most cases the equivalent widths were measured by Gaussian
fitting.  In some cases direct integration was used when the line
did not approximate well a gaussian profile.  The equivalent width
measurements are presented in Tables \ref{eq1}, \ref{eq2}, \ref{eq3}, and \ref{eq4}.

The list of atomic lines used in our abundance determinations,
including excitation potentials and $gf$ values, is given in Table
\ref{linelist}.  The list was obtained after some experimentation
using the abundance standards spectra. The initial list was based on
that used for metal-poor stars in the various HIRES studies of Johnson
and collaborators (e.g. Johnson \& Bolte 2002).  Lines in more crowded
regions were discarded because of the resolution limit of ESI. Lines
were also discarded if they consistently resulted in discrepant
abundances compared to other lines of the same element in the analysis
of the abundance standards.  Whenever the option was present, we used
experimental rather than theoretical or solar $gf$ values.  We used the
line list for the barium hyperfine splitting given in McWilliam (1998).

\section{Abundance Determinations}

\subsection{Atmospheric Parameters}
The usual spectroscopic methods for deriving atmospheric parameters in higher
resolution studies (R$>30000$) cannot be applied to the ESI spectra analysis. 
There are in general too few Fe I lines to adjust
our values of effective temperature and microturbulent velocity.
Additionally, Fe {\small II} could not be measured reliably, eliminating the option
of using ionization balance to calculate surface gravity. 

We therefore derived the atmospheric parameters for each abundance
standard using photometric values.  In almost all cases we calculated
the temperature using $B-V$ values, although when available we used the
additional temperature indicator $b-y$. As a check
we used $V-K$ derived temperatures as comparisons.  As discussed later, we
preferred $b-y$ as the temperature indicator for the
program sample.  We used the method and formulae presented in Alonso,
Arribas \& Mart\'{\i}nez-Roger (1996 and 1999) for transforming colors to
\teff. The cluster giants had colors too red to obtain accurate
temperatures via this method, and we derived abundances for these stars
only using already published atmospheric parameters.  Given effective
temperature, we then obtained surface gravity values using Bergbusch
\& Vandenberg (1992) isochrones  and an initial guess of [M/H].  
The photometry of the standards and program stars is presented in Tables
\ref{photometry} and \ref{progphot}, respectively.

Microturbulent velocity ($\xi$) was
assigned in a very simple manner.  A velocity of 2 km s$^{-1}$ was assigned
to giants (defined here by log $ g < 3.0$), 1.5 km s$^{-1}$ to subgiants ($3.0\leq
$ log $ g <3.9$) and 1 km s$^{-1}$ for main-sequence stars (log $g\geq
3.9$).  This gives a reasonable agreement with the range of $\xi$
values given in studies such as Cayrel
et al. (2004), Johnson (2002), and McWilliam et al. (1995).

Finally, the [M/H] value of the atmosphere was found by iterating on the
initial guess until the derived [Fe/H] matched the model value to
within 0.2 dex.  The formulae of Alonso
et al. (1996 and 1999) depend on
[M/H], and therefore so do the values for effective temperature and
surface gravity according to our
methods.  We iterate these atmospheric parameters until the
metallicity of the atmosphere was similar to the final stellar
metallicity.  This procedure was
carried out on all of the abundance standards and program sample.

\subsection{Reddening}

We obtained reddening estimates for $B-V$ and $V-K$ using the Schlegel et
al. (1998) dust maps.  The reddening estimates for the objects with
$b-y$ colors were taken directly from the Anthony-Twarog et al. (2000)
paper, which is based on the Burstein \& Heiles (1982) dust maps.  

For almost all of the abundance standard stars no reddening was applied
because they are relatively nearby. The exceptions were the BS/CS
abundance standards. These objects are on
average much further away than the brighter HD/BD standards.  
For this subset of the standards, the photometry used to derive 
atmospheric parameters was dereddened.
For the
program stars we calculated abundances for two cases, one assuming the
full column of dust in the direction of the stars, the other assuming
no reddening.  The atmospheres for the program stars, with and
without a reddening correction, are given in Table \ref{progatm}.

\subsection{Abundances}

We use Kurucz model atmospheres with overshooting, no
$\alpha$-enhancement, and without the new ODF, optical depth function,
(http://kurucz.harvard.edu/grids.html) and the LTE line analysis
program MOOG, written by Chris Sneden (http://verdi.as.utexas.edu/moog.html), to derive our final abundance
results.  Based on our careful selection of lines, we assumed the
lines in our final line list are unblended.

Table \ref{linedata} presents the number of lines we measured for each
element in each object, and the standard deviations of these
measurements.  We do not include Cr in this table because we only
measure its 5345.81 \AA{} line.

\subsection{Radial Velocities}

We used the FXCOR routine in IRAF to get cross-correlation-based
velocities using two abundance standard stars with velocities closely
matching previously published values, HD
63791 and HD 74462 as references.  For each program star four separate
orders were cross correlated against each velocity standard.
Obviously discordant correlation peaks were discarded and the rest averaged.
The typical RMS scatter
from these measurements was 4 km s$^{-1}$. As mentioned in \S
\ref{process} there was a correction of $\leq 5$ km s$^{-1}$ 
based on measurement of night sky lines. This correction is accurate to
$\sim 1$ km s$^{-1}$ec. There may be an
overall offset due to incorrect velocities of our `standard' objects,
but taking the above values as our random errors gives a final internal
uncertainty of $\sim 4.2$ km s$^{-1}$.

\section{Comparisons with Previous Studies: Abundance Standards}
 
\subsection{Equivalent Widths}
Figure \ref{ewfulb} compares our equivalent width values with studies
that have a significant common set of measured lines.  The overlaps are
13 stars and 70 lines for Fulbright (2000), 2 stars and 27 lines for
McWilliam et al. (1995), and 3 stars and 65 lines for Johnson (2002).
The average offsets, all in the direction of our values minus the
previous studies, are 7.3, 7.3, and 6.9 m\AA{}, for Fulbright (2000),
McWilliam et al. (1995), and Johnson (2002), respectively.  The
standard deviations, also in the same order, came out to 9.4, 12.9,
and 8.0 m\AA{}.
We believe the offset is due in part to the lower resolution
spectra from ESI.  Even with our careful selection of lines, our study
may still have blending from
additional lines not affecting these higher resolution measurements.
This effect
would make our abundances systematically too high, an effect which will
be discussed in more detail in a following section.  We did institute a
cut-off by excluding lines with measured equivalent widths greater than
200 m\AA{} in our analysis.  The best fit lines from Figure
\ref{ewfulb} does not include the few points shown to
have been measured greater than 200 m\AA{}, nor are these points
included when calculating the offset values.

\subsection{Atmospheres \label{atmsection}} 
Figure \ref{atmcompare} compares atmospheric parameters we derived from
photometric values compared to parameters given by Fulbright (1999) and
Pilachowski et al. (1996).  Overall our atmospheric parameters are
close to published values shown in Table \ref{litparam}.  Our final results are summarized in Table
\ref{photoatm}.  The standard deviation for temperatures derived from
$B-V$ temperatures is 125
K; the value is reduced to 118 K for
$b-y$ temperatures.  For this reason we used $b-y$ determined
temperatures for the program sample when it was available.  For log$g$
values determined by $B-V$ and $b-y$ derived temperatures, the
standard deviations are both 0.46 dex.  The atmospheric values determined from $V-K$
are very similar to the other colors, and the standard deviations are
127 K in temperature, and 0.54 in log$g$.

\subsection{[Fe/H]}
 
We plot $\Delta$[Fe/H] (in the sense of our values minus previous
values) versus atmospheric parameters in Figure \ref{trends1}.  
There is a $\sim 0.12$ dex offset between our values compared to 
previously published values and small trends in
log$g$, \teff and $\xi$.

Figure \ref{trends1} suggests our giant and subgiant metallicities are
approximately 0.1 dex too high, while dwarfs are an additional 0.2 dex
too high.  The overall offset follows directly from Figure
\ref{ewfulb} which show that our equivalent widths are
measured consistently too high in comparison to the higher resolution
study.  In a test case of a typical object, decreasing all of the
equivalent widths by 7.0 m\AA{} yielded a metallicity 0.1 dex lower
than the previous value, fitting the offset almost perfectly.  The
trend with surface gravity, which is also apparent in \teff and
$\xi$, is more difficult to understand, but may be a result of our simple
assumption about $\xi$. However, for the purpose of this work, to identify the
bona-fida EMP stars in the candidate lists, these accuracies for [Fe/H] are
more than sufficient.

\subsection{Other Elements}

The abundance results for the standards are presented in Tables
\ref{FeMg}, \ref{CaCr}, and \ref{TiBa}.  The final
results are also plotted in Figures \ref{litcomp}, \ref{bvcomp}, and
\ref{bycomp}.  The sense of the plots is our abundance value minus the
previously published value.

The agreement is generally within a factor of two, and roughly
consistent with errors calculated from atmosphere values.  In all of the
figures lines have been drawn at $\pm0.3$ dex.  We have
also marked the objects from the Pilachowski et al. (1996) work with
diamond symbols.  We measure the $\alpha$-element abundances too high
relative to this previous study.  Without these points, our measurements are fairly consistent
with the other previous values, but we may be measuring Ti {\small II}
too high overall.

We also carried out an average of the $\alpha$-elements for
comparisons.  This is shown in Figure \ref{alpha}.  Though there is a
noticeable scatter in the comparisons, we do quite well in reproducing
the $\alpha$ abundances to within 0.15 dex.

\section{Abundance Errors}

There are three main components to the abundance error budget. The
errors in adopted model atmosphere, uncertainties in the equivalent-width
measurements and errors in the atomic line parameters all contribute
to the final abundance uncertainty. These are discussed below.
An estimate of the combined errors can be derived from the scatter in Figures
\ref{litcomp}, \ref{bvcomp}, and \ref{bycomp} where we compare our abundance
measurements to the high-quality literature values for the 25
``standard'' stars. The RMS values of those differences give a good
estimate of the accuracy of our abundance measurements.  The $B-V$
atmosphere derived abundances compared to the abundances of previous
studies give standard deviations of 0.16, 0.22, 0.20, 0.23, 0.18,
and 0.22 dex in Fe, Mg, Ca, T{\small II}, Cr, Ba {\small II} abundances, respectively.

\subsection{Errors due to uncertainties in the adopted atmosphere parameters} 
The sensitivity of [Fe/H] and other element
abundances to each atmospheric parameter was determined by varying
each input parameter while holding the others fixed and deriving the
abundance.  Three stars were chosen for this analysis, HD 94028,
BD+37 1458, and CS 22892-052, picked to represent dwarfs, sub-giants,
and giants respectively.

We then calculated the atmospheric error assuming uncertainties of $\pm125$ K in
\teff, $\pm0.5$ dex in log$g$, $\pm0.5$ km s$^{-1}$ in $\xi$, and $\pm0.2$
in [Fe/H], and then adding the resulting [Fe/H] variations in
quadrature.  The errors for effective temperature and surface gravity
were estimated by taking the standard deviation from the differences
between the previously published values and our $B-V$ atmospheric
values.  One cross term was taken into account because of the
coupling of \teff and log$g$.  Varying the \teff by +125 K in
HD 94028, BD+37 1458, and CS 22892-052, gave changes in log$g$ of -0.04
dex, 0.17 dex, and 0.38 dex, respectively.  This effect was then added
into the quadrature sum.  Although the \teff and
log$g$ are also dependent on [M/H], the error caused by [M/H] variation
is so small that those cross terms have a negligible effect.

Tables \ref{error1},
\ref{error2}, and \ref{error3} give the values
used in this analysis and the effects on the three stars.  The final error
estimate includes the \teff and log$g$ coupling term.  For all
three stars the final errors are approximately the same, 0.2 dex for
[Fe/H], [T{\small II}/Fe], and [Ba {\small II}/Fe], and 0.1 dex for [Mg/Fe], [Ca/Fe], [Cr/Fe].

\subsection{Errors due to uncertainties in the equivalent width
measurements and atomic parameters for individual absorption lines}

When there are multiple lines measured for an element, Table
\ref{linedata} gives an estimate for the error from equivalent width and atomic
parameters.  The error for the case when only one line is measured is less
obvious.  Therefore we turn back to our abundance standard
comparisons.  To match the errors from rms scatter (an estimate of the
total error) and get answers
that agree with the literature abundances, a value of about
0.1 dex, i.e. 0.17 dex when adding in quadrature, needs to be added to the atmospheric errors in Mg, Ca, and
Cr.  We assume that this additional scatter arises from the
combination of equivalent width measurement error and incorrect atomic
parameters.  In general the $\sigma$'s from Table \ref{linedata} are
in this 0.1 to 0.15 dex range, including the T{\small II} and Ba {\small II}
measurements.  A conservative estimate for our errors would then be
our atmospheric errors plus 0.15 dex (or 0.23 dex when adding in quadrature).  This gives agreement with
general errors when multiple lines are measured, and gives
confidence in assigning this value to when only one line is used in
the abundance determination.

\section{Results}

For the program CS/BS stars we carried out the same analysis as for the
abundance standards. The final abundance results are presented in 
Tables \ref{dereddened} and \ref{reddened} for the two cases with and
without reddening correction.  
There are a number of stars with [Fe/H]$>-1.5$.  We did not include
abundance standards with [Fe/H]$>-1.5$ and make no claim of accuracy
for these `metal-rich' stars.  We can say with certainty that they are
relatively metal-rich.  The radial velocities for each of these
objects is presented in Table \ref{progatm}.  

\section{Discussion}

Figure \ref{hist} shows the abundance distribution for our
program stars. Because of poorly-characterized selection functions
that have shaped our sample, the
metallicity distribution is not meaningful. However, our goal was to identify
bona-fida EMP stars for further study, and Figure \ref{hist} shows that we have a
number of EMP stars. There are 30 stars with [Fe/H]$<-2.5$ for which 
no detailed abundance studies have been carried out, and for which
we have measured the abundances to an accuracy of about 0.3 dex. 
Surprisingly, nearly 25$\%$ of the sample turned out to be relatively
metal-rich, i.e. [Fe/H]$> -1.5$.  
 
The upper panel of Figure \ref{trends} shows the summed $\alpha$-element 
abundance ratios for the stars in our sample. Although the scatter is
large, we see that well known property of super-solar [$\alpha$/Fe] for metal-poor stars
is reproduced in these stars.  The trend towards solar [$\alpha$/Fe]
with increasing [Fe/H] is also noticeable, even though our errors may be
much larger in this regime.  The lower panel of Figure \ref{trends}
plots the Ba trend with [Fe/H].  The behavior is consistent with the
trend shown in Figure 2 in McWilliam (1998).  The Ba-rich star in
Figure \ref{trends}, CS 31062-050 has already been the subject of a recent study by
Johnson \& Bolte (2004).  We are currently obtaining and analyzing
high-resolution spectra of many of the EMP stars that we have
identified in this study.

The ESI instrument has proven capable of identifying EMP stars in
a very efficient manner down to $V\sim 14.5$.  Exposures times ranging from 10 to 20
minutes produced very good signal to noise data for the faintest of
the BS/CS sample.  From these data we are able to measure abundances
well enough to measure [Fe/H] to very low values with a precision of
$\sim 0.3$ dex, and discover low $\alpha$ stars, as well as find stars with
anamolously high Ba.

\section{Acknowledgements} 
This research is based in part upon work supported by the 
National Science Foundation under Grant Number AST-0098617
Any opinions, findings, and conclusions or recommendations 
expressed in this material are those of the author(s) and do not necessarily
reflect the views of the National Science Foundation.
The authors wish to recognize and
acknowledge the very significant cultural role and reverence that
the summit of Mauna Kea has always had within the indigenous Hawa{\small II}an
community.  We are most fortunate to have the opportunity to conduct
observations from this mountain.

\clearpage

\setcounter{figure}{1}

\begin{figure}
\begin{center}
\scalebox{.4}[.4]{
\plotone{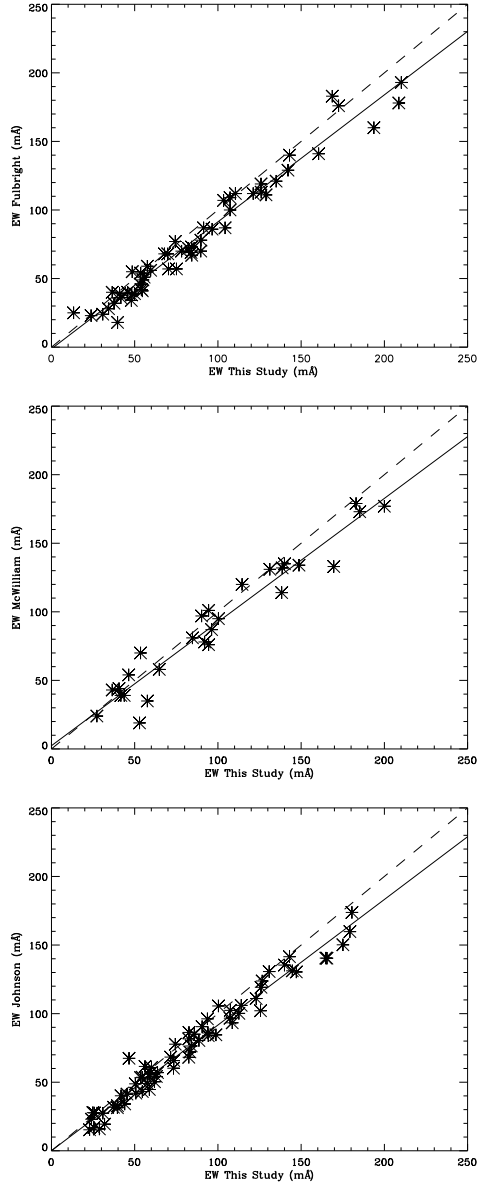}}
\end{center}
\figcaption[Lai.fig2.eps]{Equivalent width comparisons with Fulbright
et al. (2000), McWilliam et al. (1995), and Johnson (2002).
The dashed lines are the one-to-one lines, and the solid lines represent
the best fit lines.  The slopes for each best fit line are 0.93, 0.90,
and 0.92, and the intercepts are -1.2, 2.4, and 0.1 m\AA{}, in the
same order as listed before.\label{ewfulb}}
\end{figure}

\begin{figure}
\begin{center}
\scalebox{.85}[.85]{
\plotone{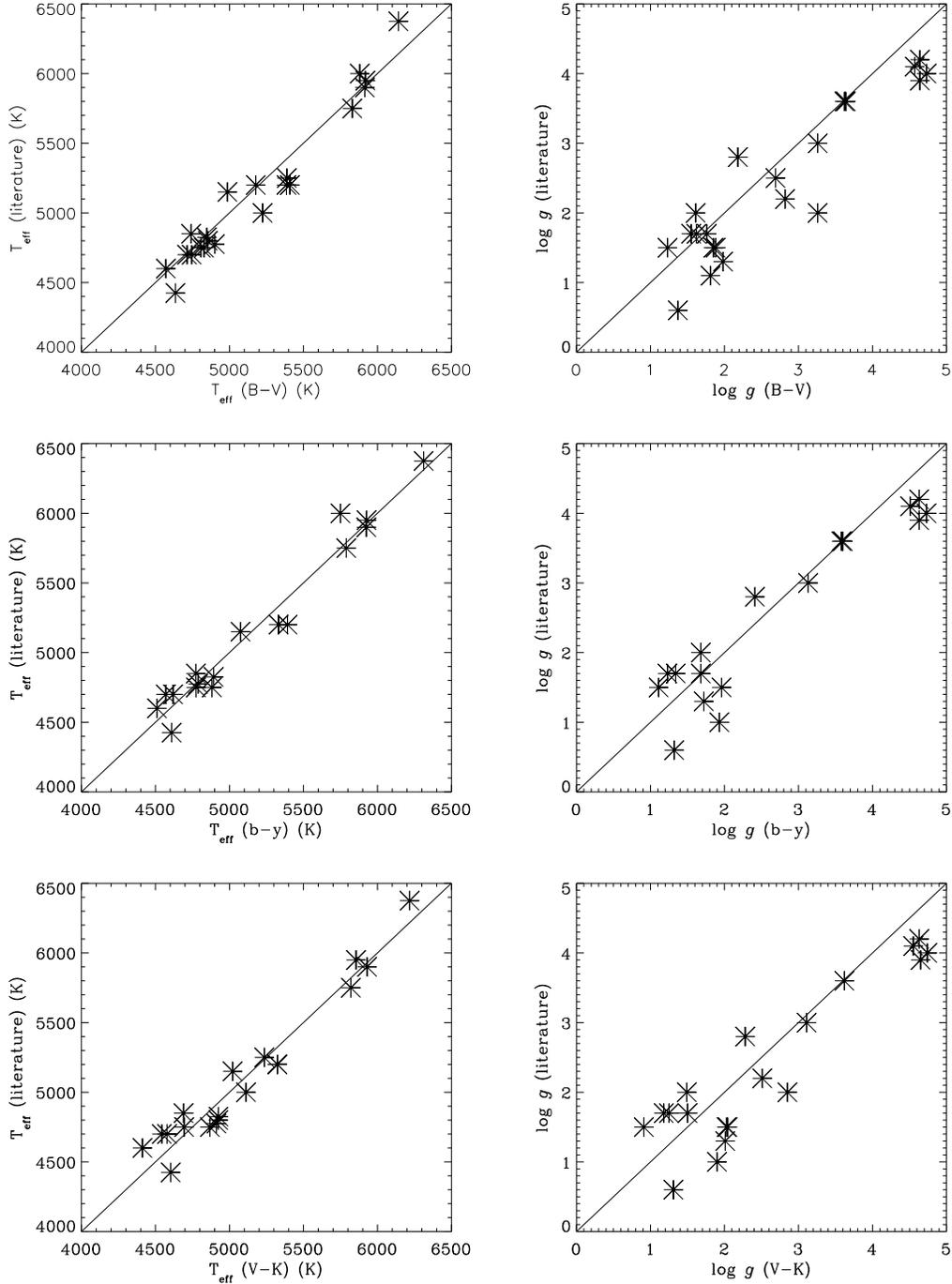}}
\end{center}
\figcaption[Lai.fig3.eps]{Atmospheric parameters comparison for the
standards, plotting our derived parameters versus previously published
parameters.  The solid
line is the one-to-one line. \label{atmcompare}}
\end{figure} 

\begin{figure}
\plotone{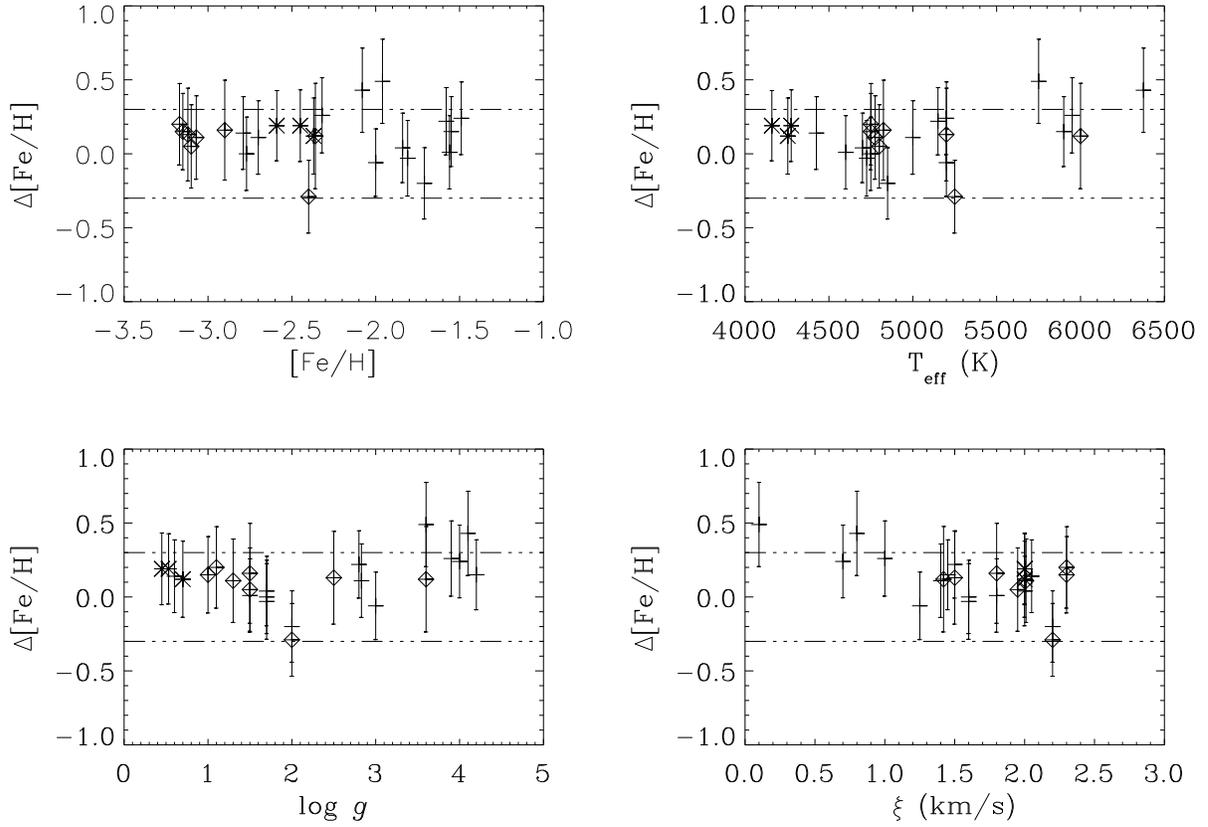}
\figcaption[Lai.fig4.eps]{A plot of atmospheric trends.  The error bars are the
errors derived from atmospheric variations ($\sim$0.2 dex) added to the
standard deviations from the line to line variations.  The diamond
symbols represent the CS/BS standard stars, the * represents the
globular cluster stars, and the crosses all others.\label{trends1}}
\end{figure} 

\begin{figure}
\plotone{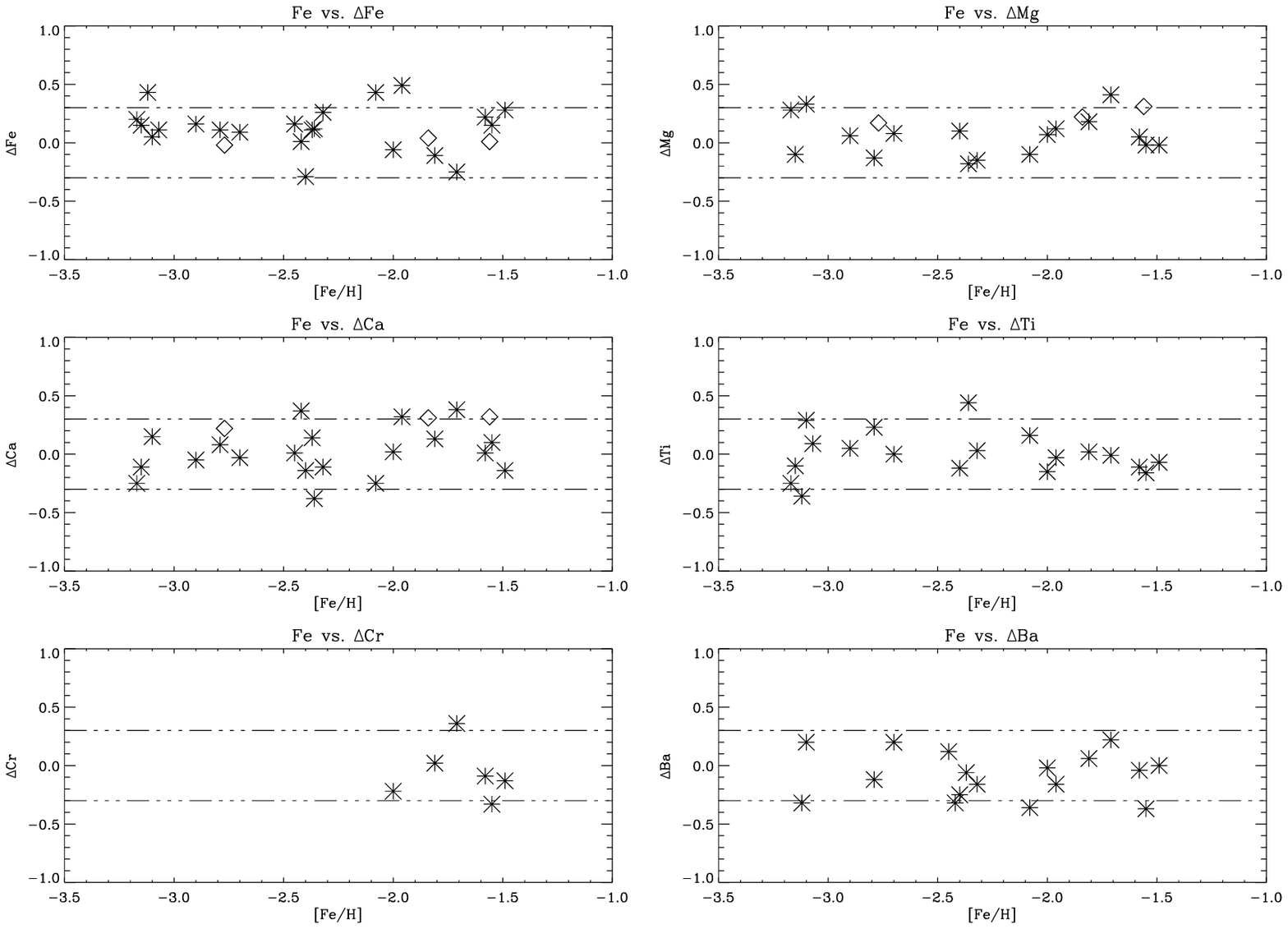}
\figcaption[Lai.fig5.eps]{Element by Element comparison using literature
atmopsheres. The diamond symbols represent Pilachowski et al.
(1996) stars, and the * all other standards. \label{litcomp}}
\end{figure} 

\begin{figure}
\plotone{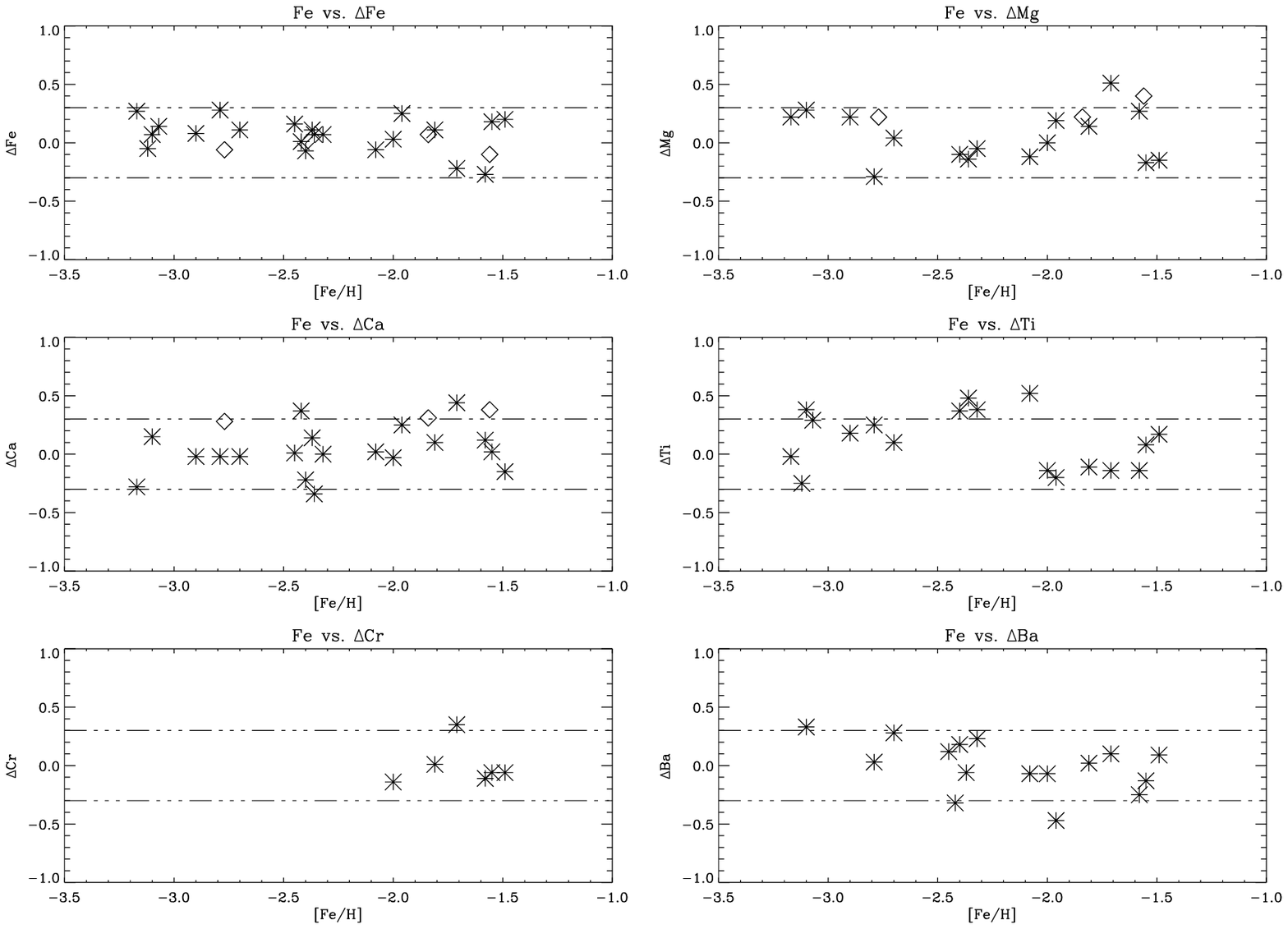}
\figcaption[Lai.fig6.eps]{Element by Element comparison using $B-V$ atmopsheres. The diamond symbols represent Pilachowski et al.
(1996) stars, and the * all other standards.\label{bvcomp}}
\end{figure} 

\begin{figure}
\plotone{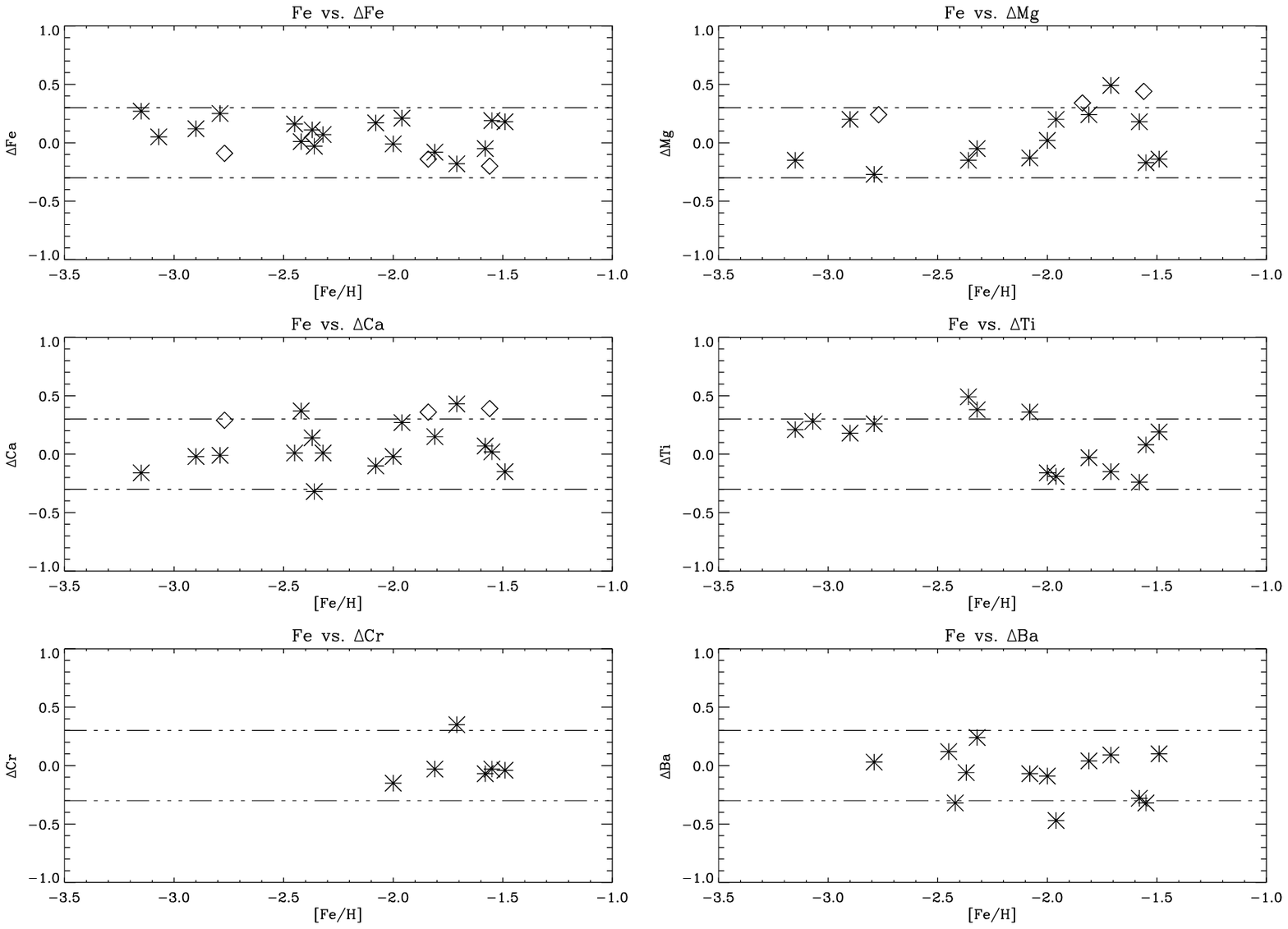}
\figcaption[Lai.fig7.eps]{Element by Element comparison using $b-y$ atmopsheres. The diamond symbols represent Pilachowski et al.
(1996) stars, and the * all other standards.\label{bycomp}}
\end{figure} 

\begin{figure}
\plotone{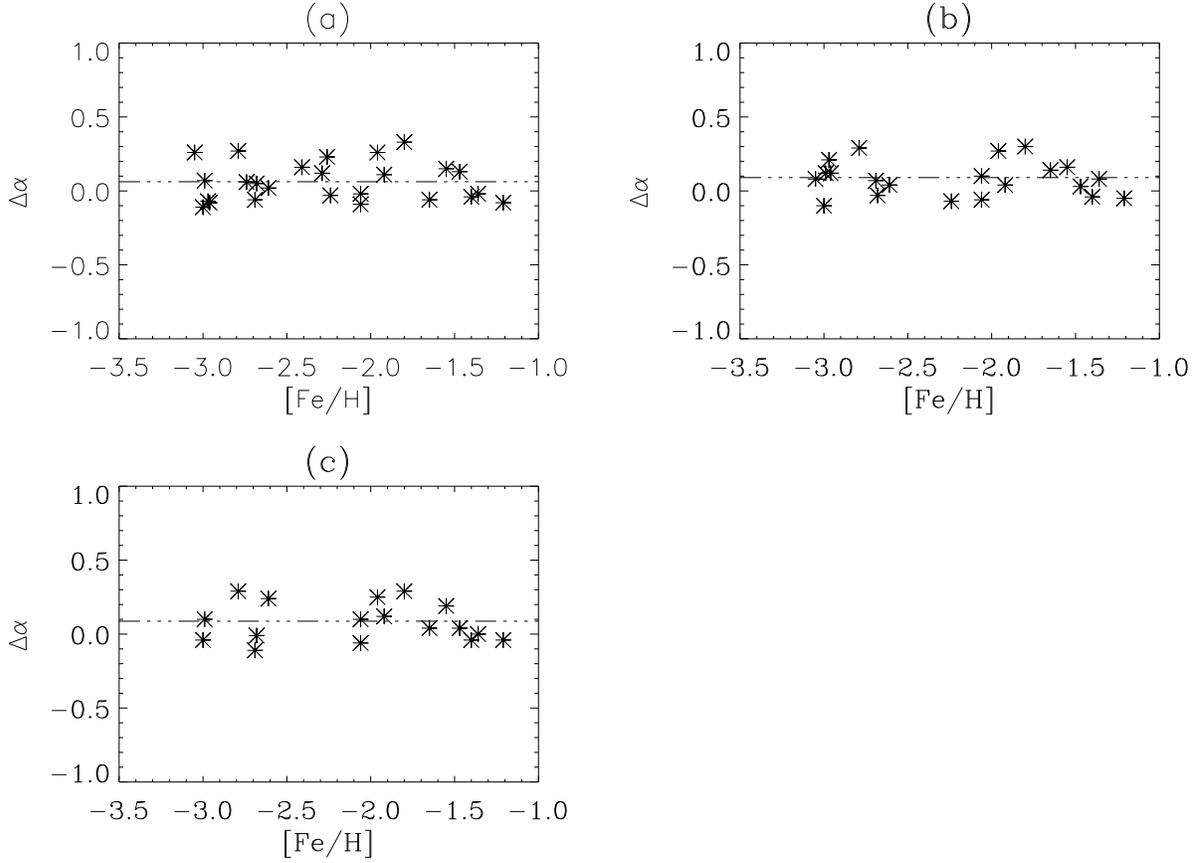}
\figcaption[Lai.fig8.eps]{Comparisons of $\alpha$ elements of the standards with
literature values.  The sense is our values minus literature.
Abundances are derived from (a) literature values, (b) $B-V$
atmospheres, and (c) $b-y$ atmospheres.  The average of the offset,
the dased lines, are 0.06, 0.09, and 0.09 dex, respectively.\label{alpha}}
\end{figure}

\begin{figure}
\plotone{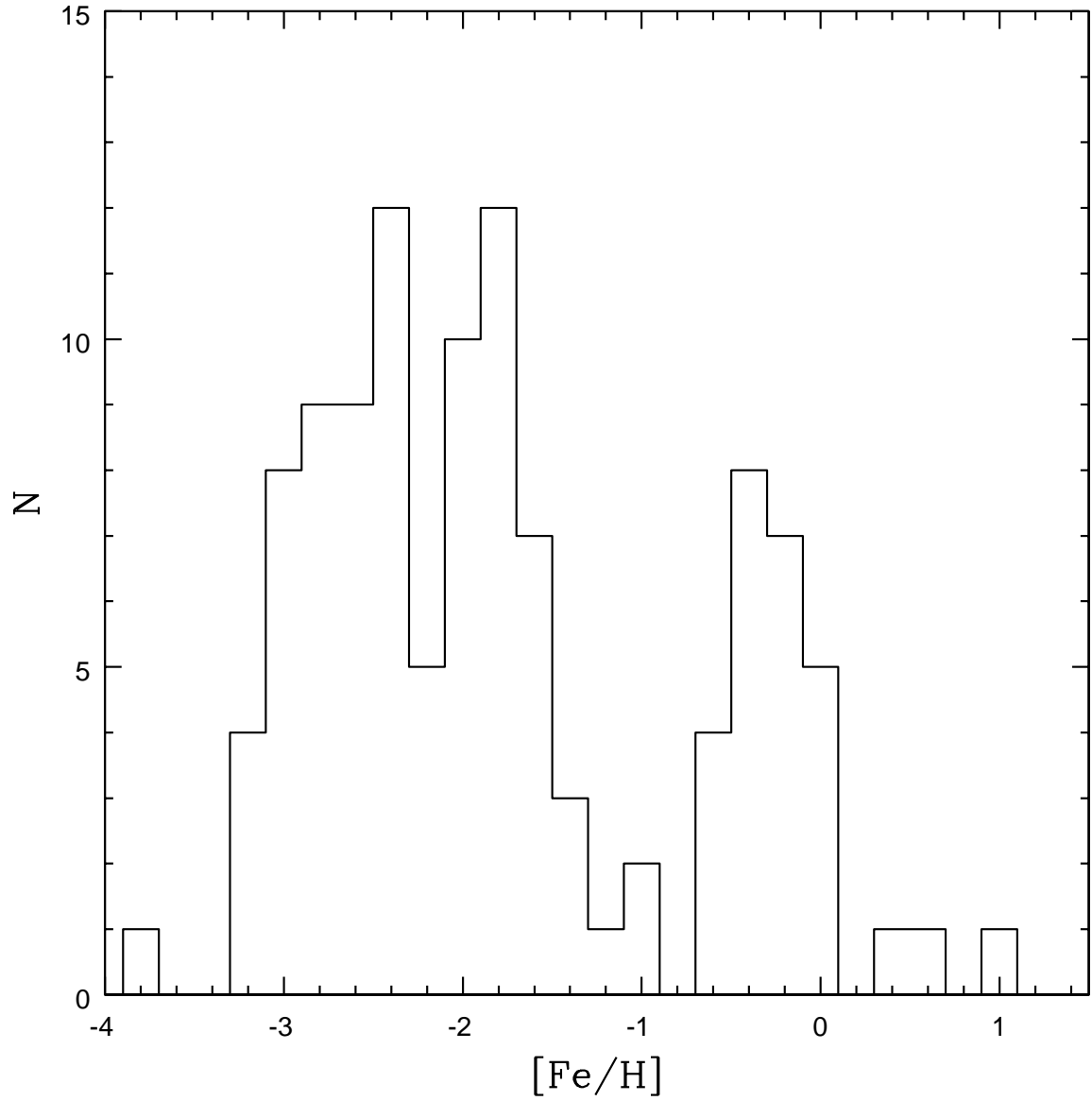}
\figcaption[Lai.fig9.eps]{The program star metallicity distribution  using bins of 0.2 dex in [Fe/H]. \label{hist}}
\end{figure} 

\begin{figure}
\plotone{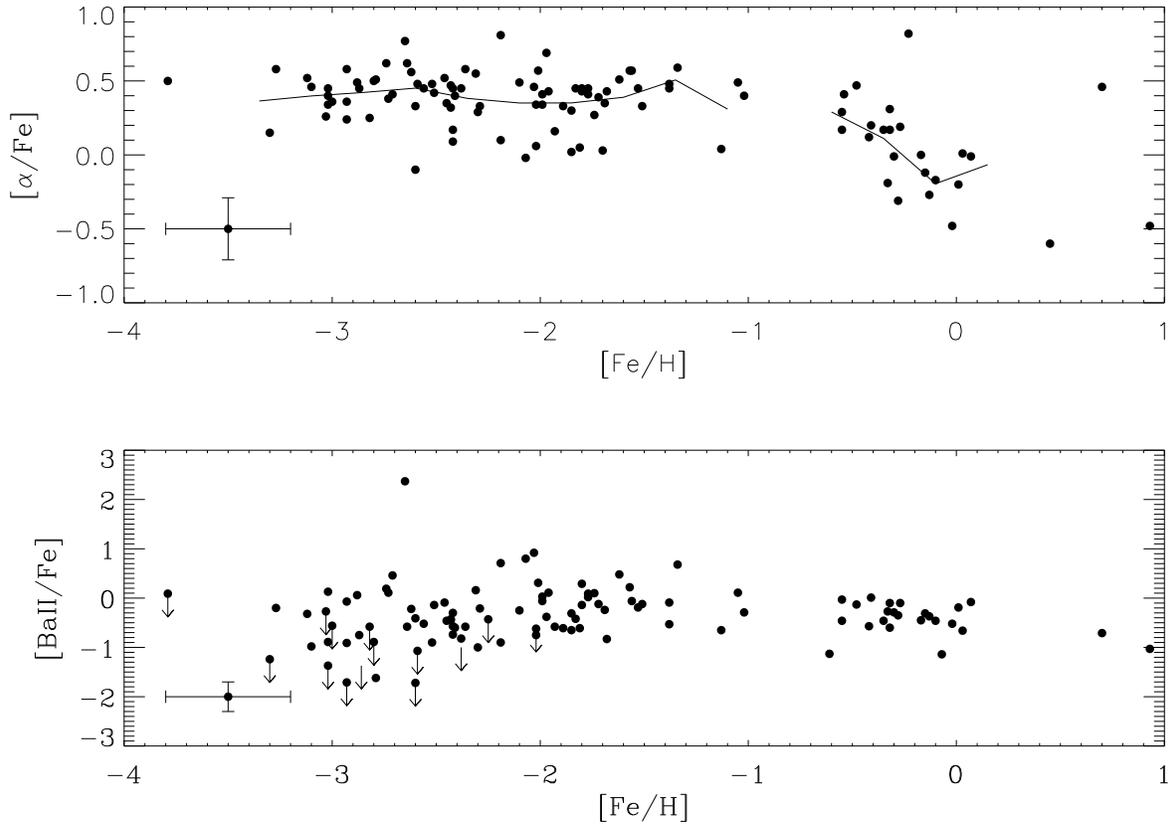}
\figcaption[Lai.fig10.eps]{Trends of averaged $\alpha$ and barium abundances of the
program stars versus
metallicity.  The arrows represent upper limits for the Ba abundances.  A representative error bar is shown in the lower
left hand corner of each plot.  A line averaging the $\alpha$ values
in bins of 0.25 dex in [Fe/H] is also included in the top plot.  \label{trends}}
\end{figure}

\begin{deluxetable}{llllccccl}
\tablecolumns{9}
\tablewidth{0pc}
\tablecaption{Photometry of Abundance Standards \label{photometry}}
\tablehead{
\colhead{Star} & \colhead{V}   & \colhead{$B-V$} & \colhead{$b-y$}
&\colhead{$V-K$}  & \colhead{Adopted} &\colhead{Phot.}
&\colhead{Exp. Time}  & \colhead{S/N} \\ \colhead{ID} &
\colhead{(mag)} & \colhead{(mag)} &  \colhead{(mag)} & \colhead{(mag)} &
\colhead{$E(B-V)$} &\colhead{Ref.} & \colhead{(s)} & \colhead{at 6035 \AA}}
\startdata
BD-03 2525 & 9.67 & 0.48 & 0.349 & 1.48 &0.00 & 1,2 & 60 & 324  \\    
BD+02 3375 & 9.93 & 0.45 & 0.352 & 1.45 &0.00& 1,2  &240 & 273  \\    
BD+23 3130 & 8.95  & 0.64 & \nodata & 2.00 &0.00& 4 &120 & 330  \\    
BD+37 1458 & 8.92 & 0.60 & 0.432 & 1.83 & 0.00 & 1,2& 60 & 245  \\    
HD 6755 & 7.73 & 0.72 & 0.487 & 2.06& 0.00 &1,2 &  8 & 170  \\    
HD 44007 & 8.06 & 0.84 & 0.559 & 2.37 & 0.00 & 1,2 & 15 & 252  \\    
HD 63791 & 7.92 & 0.85 & 0.612 & 2.49& 0.00& 2,3 & 10 & 343  \\    
HD 74462 & 8.69 & 0.97 & 0.655 & 2.69&0.00& 1,2 & 70 & 276  \\
\enddata
\tablecomments{V magnitudes taken from Simbad, and K magnitudes are taken from the Two Micron All Sky Survey.\\}
\tablerefs{(1) $B-V$ data taken from average Simbad values; (2) $b-y$
values from Hauck \& Mermilliod 1998; (3) $B-V$ from Suntzeff et
al. 1998;(4) $B-V$ from Carney \& Latham 1987; (5) $B-V$ from Sneden 1997;
(6) $B-V$ from Sneden 2000;(7) $b-y$ from Anthony-Twarog 2000;(8) $B-V$ from
Beers et al. 2000; (9) $B-V$ from Cohen et al. 2002; (10) $B-V$ from
Sneden et al. 2003;(11) $B-V$ from Hill et al. 2002}
\end{deluxetable}

\begin{deluxetable}{lcccccccccc}
\tablecolumns{11}
\tabletypesize{\scriptsize}
\setlength{\tabcolsep}{0.04in}
\tablewidth{0pc}
\tablecaption{Program Star Photometry\label{progphot}}
\tablehead{
\colhead{Star ID} & \colhead{$V$} & \colhead{$B-V$} & \colhead{$E(B-V)$} &
\colhead{$b-y$} &\colhead{$c1$} & \colhead{$(b-y)_o$} &
\colhead{$c1_o$} & \colhead{Ref.} & \colhead{Exp. time} &\colhead{Date
obs.}}
\startdata								
BS 16076-006 & 13.56 & \nodata &\nodata&0.424 &	0.730 &	0.398 &	0.725&	1  & 900 & 2003 Apr 28\\
BS 16077-007 & 12.35 & \nodata &\nodata&0.338 &	0.193 &	0.338 &	0.193&	1  & 400 & 2001 jan 21\\
BS 16080-054 & 12.79 & \nodata &\nodata&0.564 &	0.474 &	0.562 &	0.474&	1  & 450 & 2001 Jul 7\\
BS 16080-093 & 13.61 &	0.69  &	0.02  &\nodata&\nodata&\nodata&\nodata&	2  & 600 & 2001 Jul 7\\
BS 16084-160 & 13.16 & \nodata&\nodata&	0.615 &	0.473 &	0.610 &	0.472&	1  & 450 & 2001 Jul 7\\
BS 16467-062 & 14.03 & \nodata&\nodata&	0.480 &	0.222 &	0.480 &	0.222&	1  & 900 & 2003 Apr 28\\
BS 16470-073 & 14.16 &	0.73  &	0.04  &\nodata&\nodata&\nodata&\nodata&	3  & 900 & 2003 Apr 23\\
BS 16470-092 & 13.94 &	0.48  &	0.03  &\nodata&\nodata&\nodata&\nodata&	3  &1200 & 2003 Apr 23\\
\enddata	
\tablenotetext{\dag}{The photometry information from Allende Prieto et al. 2000 was derived directly into dereddened colors from R $\simeq$ 2000 spectra, and therefore we do not apply a reddening correction to any of these stars.}
\tablenotetext{*}{ These spectra were taken with a 1\Sec25-wide slit, and
therefore the standard calibrations do not fully apply to them.}
\tablerefs{(1) Anthony-Twarog et al. 2000; (2) Bonifacio et al. 2000;
(3) Norris et al. 1997;(4) Allende Prieto et al. 2000; (5) Beers et
al. 2000}
\end{deluxetable}			
\clearpage

\begin{deluxetable}{llllllllllll}
\tablecolumns{12}
\tabletypesize{\scriptsize}
\setlength{\tabcolsep}{0.03in}
\tablecaption{EW measurements of Mg and Ca in m\AA{}\label{eq1}}
\tablehead{
\colhead{Star} & \colhead{Mg{\small I}} & \colhead{Mg{\small I}} & \colhead{Mg{\small I}}
&\colhead{Mg{\small I}} & \colhead{Ca{\small I}} & \colhead{Ca{\small I}} &\colhead{Ca{\small I}} & \colhead{Ca{\small I}} & \colhead{Ca{\small I}} & \colhead{Ca{\small I}} &\colhead{Ca{\small I}} \\ \colhead{ID} &\colhead{4167.27} &\colhead{5172.69} &\colhead{5183.61} &\colhead{5528.41} &\colhead{4425.44} &\colhead{5586.76} &\colhead{6122.22} &\colhead{6162.18} &\colhead{6439.08} &\colhead{6471.66} &\colhead{6717.68}} 
\startdata
BD-03 2525  &\nodata&\nodata&\nodata&100.2&73.2&\nodata&82.1&93.1&69.2&\nodata&\nodata                \\
BD+02 3375  &45.1&\nodata&\nodata&61.6&35.0&\nodata&38.5&57.2&45.2&\nodata&\nodata                    \\
BD+23 3130  &\nodata&\nodata&\nodata&70.3&38.9&42.1&54.7&70.5&52.6&\nodata&\nodata                    \\
BD+37 1458  &\nodata&\nodata&\nodata&96.5&67.9&\nodata&74.0&81.9&70.5&\nodata&\nodata                 \\
HD 6755     &\nodata&\nodata&\nodata&128.9&86.1&\nodata&125.3&129.1&113.3&\nodata&50.4                \\
HD 44007    &\nodata&\nodata&\nodata&142.1&121.7&\nodata&120.9&166.3&109.2&65.4&66.7                  \\
HD 63791    &\nodata&\nodata&\nodata&143.0&114.0&\nodata&130.8&165.5&126.6&71.6&64.7                  \\
HD 74462    &\nodata&\nodata&\nodata&166.8&141.9&\nodata&142.5&\nodata&136.0&\nodata&86.0             \\
\enddata				     	      						      
\end{deluxetable}										      
\clearpage												      

\begin{deluxetable}{lllllllllllll}
\tablecolumns{13}
\tabletypesize{\scriptsize}
\setlength{\tabcolsep}{0.02in}
\tablecaption{EW measurements of Ti {\small II},Cr, and Ba {\small II} in m\AA{}\label{eq2}}
\tablehead{
\colhead{Star} &\colhead{Ti{\small II}} &\colhead{Ti{\small II}} & \colhead{Ti{\small II}} &\colhead{Ti{\small II}} & \colhead{Ti{\small II}} & \colhead{Ti{\small II}} &\colhead{Ti{\small II}} & \colhead{Ti{\small II}} &\colhead{Ti{\small II}} & \colhead{Cr{\small I}} &\colhead{Ba{\small II}} &\colhead{Ba{\small II}}\\ \colhead{ID}& \colhead{4028.34} &\colhead{4395.04} &\colhead{4399.77} &\colhead{4443.81} &\colhead{4468.50} &\colhead{4563.76} &\colhead{4571.98} &\colhead{4589.95} &\colhead{5336.79} &\colhead{5345.80} &\colhead{4554.04} &\colhead{6141.75}}
\startdata
BD-03 2525  &\nodata&92.1&\nodata&\nodata&\nodata&75.8&\nodata&27.5&\nodata&\nodata&82.3&35.4\\                           
BD+02 3375  &\nodata&83.6&42.0&\nodata&68.2&60.0&62.3&\nodata&\nodata&\nodata&49.9&\nodata\\                             
BD+23 3130  &\nodata&128.2&\nodata&\nodata&\nodata&75.7&\nodata&39.7&\nodata&\nodata&89.8&30.8\\                         
BD+37 1458  &28.0&\nodata&\nodata&\nodata&\nodata&76.1&\nodata&45.3&23.2&22.8&107.3&75.1\\                               
HD 6755     &70.0&\nodata&\nodata&\nodata&\nodata&121.6&\nodata&73.8&66.9&73.0&160.5&104.3\\                             
HD 44007    &76.7&\nodata&\nodata&\nodata&\nodata&135.9&\nodata&88.9&74.8&88.2&172.4&126.0\\                             
HD 63791    &\nodata&\nodata&\nodata&\nodata&\nodata&159.8&\nodata&82.6&94.4&90.5&180.5&125.9\\                          
HD 74462    &\nodata&\nodata&\nodata&\nodata&\nodata&\nodata&\nodata&98.8&92.4&110.5&147.5&\nodata\\
\enddata				     	      						      
\end{deluxetable}

\begin{deluxetable}{lllllllllllllll}
\tablecolumns{15}
\tabletypesize{\scriptsize}
\setlength{\tabcolsep}{0.018in}
\tablecaption{EW measurements of Fe in m\AA{} \label{eq3}}
\tablehead{
\colhead{Star} &\colhead{Fe{\small I}} &\colhead{Fe{\small I}} & \colhead{Fe{\small I}} &\colhead{Fe{\small I}} & \colhead{Fe{\small I}} & \colhead{Fe{\small I}} &\colhead{Fe{\small I}} & \colhead{Fe{\small I}} &\colhead{Fe{\small I}} & \colhead{Fe{\small I}} &\colhead{Fe{\small I}} &\colhead{Fe{\small I}} &\colhead{Fe{\small I}} &\colhead{Fe{\small I}}\\ \colhead{ID}& \colhead{4045.82} &\colhead{4071.74} &\colhead{4383.55} &\colhead{4404.76} &\colhead{4442.35} &\colhead{4919.00} &\colhead{5068.77} &\colhead{5074.75} &\colhead{5133.69} &\colhead{5216.28} &\colhead{5232.95} &\colhead{5302.30} &\colhead{5324.19} &\colhead{5383.38}}
\startdata
BD-03 2525 &\nodata&\nodata&\nodata&195.2&\nodata&\nodata&39.7&\nodata&\nodata&\nodata&99.1&\nodata&55.3&\nodata\\     
BD+02 3375 &150.6&134.2&129.0&112.6&\nodata&36.5&\nodata&\nodata&22.9&\nodata&55.0&\nodata&38.1&33.4\\
BD+23 3130 &\nodata&135.6&154.4&146.6&\nodata&57.6&23.1&\nodata&\nodata&\nodata&67.9&26.3&51.4&\nodata\\
BD+37 1458 &\nodata &187.4 &\nodata &196.3 &\nodata &84.2 &37.6 &26.3 &42.2 &\nodata &91.9 &35.8 &67.3 &39.7\\
HD 6755    &\nodata &\nodata &\nodata &\nodata &\nodata &\nodata &89.7 &68.5 &69.9 &\nodata &133.0 &88.1 &113.0 &78.8\\
HD 44007   &\nodata &\nodata &\nodata &\nodata &\nodata &\nodata &83.7 &53.7 &69.8 &\nodata &133.2 &86.1 &104.2 &77.8\\
HD 63791   &\nodata &\nodata &\nodata &\nodata &\nodata &\nodata &94.0 &74.4 &81.4 &\nodata &144.7 &94.2 &122.2 &98.8\\
HD 74462   &\nodata &\nodata &\nodata &\nodata &\nodata &\nodata &122.0 &\nodata &104.4 &\nodata &144.2 &\nodata &134.7 &\nodata\\
\enddata				     	      						      
\end{deluxetable}

\begin{deluxetable}{llllllllllllll}
\tablecolumns{14}
\tabletypesize{\scriptsize}
\setlength{\tabcolsep}{0.021in}
\tablecaption{EW measurements of Fe in m\AA{} \label{eq4}}
\tablehead{
\colhead{Star} &\colhead{Fe{\small I}} &\colhead{Fe{\small I}} & \colhead{Fe{\small I}} &\colhead{Fe{\small I}} & \colhead{Fe{\small I}} & \colhead{Fe{\small I}} &\colhead{Fe{\small I}} &\colhead{Fe{\small I}} & \colhead{Fe{\small I}} &\colhead{Fe{\small I}} &\colhead{Fe{\small I}} &\colhead{Fe{\small I}} &\colhead{Fe{\small I}}\\ \colhead{ID}& \colhead{5393.17} &\colhead{5405.78} &\colhead{5434.53} &\colhead{5586.77} &\colhead{6065.49} &\colhead{6230.73} &\colhead{6252.56} &\colhead{6393.61} &\colhead{6411.65} &\colhead{6430.85} &\colhead{6609.11} &\colhead{6677.99} &\colhead{6750.15}}
\startdata
BD-03 2525 &\nodata&\nodata&63.9&\nodata&\nodata&\nodata&28.2&\nodata&\nodata&\nodata&\nodata&\nodata&\nodata\\        
BD+02 3375 &16.3&\nodata&52.0&\nodata&\nodata&\nodata&\nodata&\nodata&\nodata&\nodata&\nodata&\nodata&\nodata\\
BD+23 3130 &22.9&90.1&87.9&45.8&23.8&48.2&\nodata&\nodata&\nodata&13.3&\nodata&33.3&\nodata\\    
BD+37 1458 &34.7 &98.6 &84.0 &70.8 &41.7 &48.5 &45.8 &53.6 &\nodata &40.9 &\nodata&48.7 &\nodata\\       
HD 6755    &79.1 &\nodata &126.0 &\nodata &84.7 &102.6 &84.8 &90.9 &67.8 &82.9 &\nodata &82.6 &48.0\\       
HD 44007   &79.7 &\nodata &140.6 &\nodata &90.7 &102.7 &\nodata &89.4 &69.7 &88.2 &\nodata &95.9 &43.6\\
HD 63791   &104.3 &179.3 &158.4 &\nodata &93.7 &134.2 &\nodata &100.3 &83.2 &112.6 &\nodata &125.6 &\nodata\\
HD 74462   &106.1 &\nodata &\nodata &\nodata &109.9 &144.9 &115.4 &138.4 &100.6 &121.5 &73.7 &137.3 &71.8\\
\enddata				     	      						      
\end{deluxetable}

\begin{deluxetable}{lrrrrr}
\tablecolumns{5}
\tabletypesize{\footnotesize}
\tablewidth{5.5in}
\tablecaption{Excitation potentials and oscillator strengths \label{linelist}}
\tablehead{
\colhead{} & \colhead{$\lambda$}   & \colhead{}   & \colhead{EP}
& \colhead{}\\ \colhead{Species} &
\colhead{(\AA{})} & \colhead{log $gf$}
& \colhead{(eV)} & \colhead{Source}}
\startdata
Mg{\small I} &	 4167.27 &  -0.770 &  4.34 &  1 \\
Mg{\small I} &	 5172.69 &  -0.381 &  2.71 &  2 \\
Mg{\small I} &	 5183.61 &  -0.158 &  2.72 &  2 \\
Mg{\small I} &	 5528.41 &  -0.480 &  4.34 &  3 \\
Ca{\small I} &	 4425.44 &  -0.358 &  1.88 &  4 \\
Ca{\small I} &	 5588.76 &   0.360 &  2.51 &  5 \\
Ca{\small I} &	 6122.22 &  -0.409 &  1.89 &  6 \\
Ca{\small I} &	 6162.18 &   0.100 &  1.90 &  6 \\
\enddata				     	     
\tablerefs{(1)J\"{o}nsson et al. 1984; (2) Carretta et al.
2002; (3) Bragaglia et al. 2001; (4) Wiese \& Martin 1980; (5)
Smith \& Raggett 1981; (6) VALD database (as of 2004 May), Kupka et al. 1999; (7) Magain
1990, Zhao \& Magain 1990; (8) Bizzarri et al. 1993; (9) Martin
et al. 1998; (10) Blackwell et al. 1984; (11) Blackwell
et al. 1980; (12) Fuhr et al. 1998; (13) Blackwell {\it et
al.} 1982; (14) O'Brian et al. 1991; (15) Bard et al.
1991; (16) Blackwell et al. 1979; (17) Clementini et al.
2000}
\end{deluxetable}
\clearpage

\begin{deluxetable}{llllllll}
\tablecolumns{8}
\tabletypesize{\footnotesize}
\tablewidth{0pc}
\tablecaption{Program Star Atmospheres and Radial Velocities\label{progatm}}
\tablehead{
\colhead{Star} & \colhead{T$_{eff}$} & \colhead{log$g$} &
\colhead{[M/H]} &\colhead{T$_{effo}$} & \colhead{log$g_o$} &
\colhead{[M/H]$_o$} &\colhead{V$_{rad}$}\\ \colhead{ID} & \colhead{(K)}
&\colhead{(dex)} & \colhead{(dex)} & \colhead{(K)}&\colhead{(dex)} &
\colhead{(dex)} & \colhead{(km s$^{-1}$)}}
\startdata				
BS 16076-006 &	5458 &  3.37 &  -3.00 &	5614 &	3.51 &	-3.00	 &  202\\
BS 16077-007 &	6008 &	3.70 &	-3.00 &	5958 &	3.68 &	-2.75	 &  -30\\
BS 16080-054 &	4832 &	1.82 &	-2.75 & 4838 &	1.83 &	-2.75	 & -124\\
BS 16080-093 &	5059 &	2.37 & 	-2.75 &	5120 &	2.53 &	-2.75	 & -196\\
BS 16084-160 &	4751 &	1.63 & 	-3.00 &	4762 &	1.66 &	-3.00	 & -108\\
BS 16467-062 &	5219 & 	2.80 &	-3.50 &	5219 & 	2.80 &	-3.50	 &  -88\\
BS 16470-073 &	5134 &	2.57 &	-0.75 &	5309 &	3.07 &	-0.75	 &   24\\
BS 16470-092 &	5812 &	3.61 &	-2.00 &	5948 &	3.67 &	-1.75	 &   28\\
\enddata
\tablecomments{The $o$ subscript designates the dereddened parameters.}	
\tablenotetext{\dag}{These objects have derived photometry from Allende Prieto et al. 2000, where only intrinsic $B-V_o$ color is given.}   
\tablenotetext{*}{The photometry without reddening correction falls out of our range to derive an atmosphere.}
\end{deluxetable}
\clearpage

\begin{deluxetable}{lrrrrrrrrrr}
\tablecolumns{11}
\tabletypesize{\footnotesize}
\tablewidth{0pc}
\tablecaption{Line Information\label{linedata}}
\tablehead{
\colhead{Star} & \colhead{Fe} & \colhead{$\sigma$} & \colhead{Mg}
&\colhead{$\sigma$} & \colhead{Ca} & \colhead{$\sigma$} &
\colhead{Ti{\small II}} & \colhead{$\sigma$} & \colhead{Ba{\small II}} &
\colhead{$\sigma$} \\ \colhead{ID} & \colhead{lines} &
\colhead{Fe} & \colhead{lines} & \colhead{Mg} & \colhead{lines} &
\colhead{Ca} & \colhead{lines} & \colhead{Ti{\small II}} & \colhead{lines} &
\colhead{Ba{\small II}}}
\startdata
BD-03 2525  &	6   &	0.31 &	1      &\nodata&4      &0.21   &3      &0.12   &2      &0.23	 \\
BD+02 3375  &	11  &	0.19 &	2      &0.02   &4      &0.07   &5      &0.10   &1      &\nodata	 \\
BD+23 3130  &	16  &	0.15 &	1      &\nodata&5      &0.11   &3      &0.36   &2      &0.27	 \\
BD+37 1458  &	20  & 	0.12 &	1      &\nodata&4      &0.16   &4      &0.16   &2      &0.06	 \\
HD 6755     &	17  &	0.17 &	1      &\nodata&5      &0.21   &4      &0.21   &2      &0.24	 \\
HD 44007    &	16  & 	0.12 &  1      &\nodata&6      &0.22   &4      &0.11   &2      &0.13	 \\
HD 63791    & 	16  &	0.19 &	1      &\nodata&6      &0.07   &3      &0.38   &2      &0.22	 \\
HD 74462    &	14  &	0.19 &	1      &\nodata&4      &0.24   &2      &0.06   &1      &\nodata	 \\
\enddata				     	      
\end{deluxetable}
\clearpage

\begin{deluxetable}{llllll}
\tablecolumns{6}
\tablewidth{0pc}
\tablecaption{Stellar Parameters from Previous Studies \label{litparam}}
\tablehead{
\colhead{Star} & \colhead{T$_{eff}$}   & \colhead{log $g$}   & \colhead{[Fe/H]}
& \colhead{$\xi$} & \colhead{Ref.}\\ \colhead{ID} & \colhead{(K)} & \colhead{(dex)}
& \colhead{(dex)} & \colhead{(km s$^{-1}$)} & \colhead{}}
\startdata
BD-03 2525    &	5750   &	3.6  &    -1.96 &  0.10 & 1\\
BD+02 3375    &	5950   &	3.9  &    -2.32 &  1.00 & 1\\
BD+23 3130    &	5000   &	2.8  &    -2.70 &  1.40 & 1\\
BD+37 1458    &	5200   &	3.0  &    -2.00 &  1.25	& 1\\
HD 6755       &	5150   &	2.8  &    -1.58 &  1.50 & 1\\
HD 44007      &	4850   &	2.0  &    -1.71 &  2.20 & 1\\
HD 63791      &	4725   &	1.7  &    -1.81 &  1.60 & 1\\
HD 74462      &	4600   &	1.5  &    -1.56 &  1.80 & 2\\
\enddata
\tablerefs{(1) Fulbright 2000; (2) Pilachowski et al. 1996; (3) Kraft
\& Ivans 2003; (4) Sneden et al. 1997; (5) Sneden et al. 2000; (6)
Giridhar et al. 2001; (7) Johnson \& Bolte 2002; (8) Carretta et al. 2002; (9) Sneden et al. 2003; (10) Aoki et al. 2002;
(11) McWilliam et al. 1995; (12) McWilliam et al. 1998; (13) Hill et
al. 2002}					     
\end{deluxetable}
\clearpage
  
\begin{deluxetable}{lllllllll}
\tablecolumns{7}
\tablewidth{0pc}
%\rotate
\tablecaption{Derived Stellar Parameters and Measured Radial Velocities\label{photoatm}}
\tablehead{
\colhead{Star} & \colhead{T$_{eff}$}   & \colhead{log $g$}   & \colhead{T$_{eff}$}
& \colhead{log $g$} & \colhead{T$_{eff}$} &\colhead{log $g$} & \colhead{[Fe/H]} 
& \colhead{V$_{rad}$}\\ \colhead{ID} & \colhead{($B-V$)} & \colhead{($B-V$)}
& \colhead{($b-y$)} & \colhead{($b-y$)} & \colhead{($V-K$)} & \colhead{($V-K$)}
& \colhead{(dex)} & \colhead{(km s$^{-1}$)}}
\startdata
BD-03 2525 & 5831 & 3.62 & 5789 & 3.60 & 5820 & 3.62 & -1.75 & 33\\
BD+02 3375 & 5918 & 4.64 & 5926 & 4.63 & 5855 & 4.65 & -2.25 & -402\\
BD+23 3130 & 5224 & 2.82 & \nodata & \nodata & 5112 & 2.51 &-2.50 & -290\\
BD+37 1458 & 5386 & 3.26 & 5332 & 3.13 & 5324 & 3.11 & -2.00 & 241\\
HD 6755 & 4986 & 2.18 & 5075 & 2.41 & 5022 & 2.28 &-1.50 & -310\\
HD 44007 & 4741 & 1.61 & 4773 & 1.68 & 4690 & 1.49 &-2.00 & 167\\
HD 63791 & 4745 & 1.62 & 4619 & 1.34 & 4578 & 1.25 &-1.75 & -109\\
HD 74462 & 4571 & 1.24 & 4510 & 1.11 & 4412 & 0.91 & -1.75 & -167\\
\enddata					     
\end{deluxetable}
\clearpage

\begin{deluxetable}{lcccccccc}
\tablecolumns{9}
\tablewidth{0pc}
%\rotate
\tabletypesize{\footnotesize}
\tablecaption{[Fe/H] and [Mg/Fe] for abundance standards\label{FeMg}}
\tablehead{
\colhead{Star} & \colhead{[Fe/H]}  & \colhead{[Fe/H]}   & \colhead{[Fe/H]}
& \colhead{[Fe/H]} & \colhead{[Mg/Fe]} &\colhead{[Mg/Fe]}
&\colhead{[Mg/Fe]} &\colhead{[Mg/Fe]}\\
\colhead{ID} & \colhead{previous} & \colhead{lit. atm.} & \colhead{($B-V$)}
& \colhead{($b-y$)} & \colhead{previous} & \colhead{lit. atm.} & \colhead{($B-V$)}
& \colhead{($b-y$)}}
\startdata
BD-03 2525  & -1.96 & -1.47 & -1.71 & -1.75 & 0.37 & 0.49 & 0.56 & 0.57\\
BD+02 3375  & -2.32 & -2.06 & -2.25 & -2.25 & 0.52 & 0.37 & 0.47 & 0.47\\
BD+23 3130  & -2.70 & -2.61 & -2.59 & \nodata& 0.55 & 0.63 & 0.59 & \nodata\\
BD+37 1458  & -2.00 & -2.06 & -1.97 & -2.01 & 0.54 & 0.61 & 0.54 & 0.56\\
HD 6755     & -1.58 & -1.36 & -1.85 & -1.63 & 0.26 & 0.31 & 0.53 & 0.44\\
HD 44007    & -1.71 & -1.96 & -1.93 & -1.89 & 0.33 & 0.74 & 0.84 & 0.82\\
HD 63791    & -1.81 & -1.92 & -1.70 & -1.87 & 0.43 & 0.61 & 0.57 & 0.67\\
HD 74462    & -1.56 & -1.55 & -1.66 & -1.76 & 0.38 & 0.69 & 0.78 & 0.82\\
\enddata					     
\end{deluxetable}
\clearpage

\begin{deluxetable}{lcccccccc}
\tablecolumns{9}
%\rotate
\tabletypesize{\footnotesize}
\tablewidth{0pc}
\tablecaption{[Ca/Fe] and [Cr/Fe] for abundance standards\label{CaCr} }
\tablehead{
\colhead{Star} & \colhead{[Ca/Fe]}  & \colhead{[Ca/Fe]}   & \colhead{[Ca/Fe]}
& \colhead{[Ca/Fe]} & \colhead{[Cr/Fe]} &\colhead{[Cr/Fe]}
&\colhead{[Cr/Fe]} &\colhead{[Cr/Fe]}\\
\colhead{ID} & \colhead{previous} & \colhead{lit. atm.} & \colhead{($B-V$)}
& \colhead{($b-y$)} & \colhead{previous} & \colhead{lit. atm.} & \colhead{($B-V$)}
& \colhead{($b-y$)}}
\startdata
BD-03 2525  & 0.24 & 0.56 & 0.42 & 0.44 & -0.14 & \nodata & \nodata & \nodata\\
BD+02 3375  & 0.38 & 0.27 & 0.38 & 0.39 & 0.00 & \nodata & \nodata & \nodata\\
BD+23 3130  & 0.40 & 0.37 & 0.38 & \nodata& -0.17 & \nodata & \nodata & \nodata\\
BD+37 1458  & 0.38 & 0.40 & 0.35 & 0.36 & -0.17 & -0.39 & -0.31 & -0.32\\
HD 6755     & 0.27 & 0.28 & 0.39 & 0.34 & -0.07 & -0.16 & -0.18 & -0.14\\
HD 44007    & 0.28 & 0.66 & 0.72 & 0.71 & -0.29 & 0.07 & 0.06 & 0.06\\
HD 63791    & 0.40 & 0.53 & 0.50 & 0.55 & -0.20 & -0.18 & -0.19 & -0.23\\
HD 74462    & 0.19 & 0.51 & 0.57 & 0.58 & \nodata & -0.19 & -0.17 & -0.22\\
\enddata					     
\end{deluxetable}
\clearpage

\begin{deluxetable}{lcccccccc}
%\rotate
\tabletypesize{\footnotesize}
\tablecolumns{9}
\tablewidth{0pc}
\tablecaption{[T{\small II}/Fe] and [Ba {\small II}/Fe] for abundance standards\label{TiBa}}
\tablehead{
\colhead{Star} & \colhead{[T{\small II}/Fe]}  & \colhead{[T{\small II}/Fe]}   & \colhead{[T{\small II}/Fe]}
& \colhead{[T{\small II}/Fe]} & \colhead{[Ba {\small II}/Fe]} &\colhead{[Ba {\small II}/Fe]}
&\colhead{[Ba {\small II}/Fe]} &\colhead{[Ba {\small II}/Fe]}\\
\colhead{ID} & \colhead{previous} & \colhead{lit. atm.} & \colhead{($B-V$)}
& \colhead{($b-y$)} & \colhead{previous} & \colhead{lit. atm.} & \colhead{($B-V$)}
& \colhead{($b-y$)}}
\startdata
BD-03 2525  & 0.25 & 0.22 & -0.02 & -0.01 & 0.10 & -0.06 & -0.37 & -0.37\\
BD+02 3375  & 0.42 & 0.45 & 0.80 & 0.80 & -0.25 & -0.41 & -0.02 & -0.01\\
BD+23 3130  & 0.23 & 0.23 & 0.33 & \nodata& -0.53 & -0.33 & -0.25 & \nodata\\
BD+37 1458  & 0.18 & 0.03 & 0.04 & 0.02 & 0.09 & 0.07 & 0.02 & 0.00\\
HD 6755     & 0.24 & 0.13 & 0.10 & 0.00 & 0.03 & -0.01 & -0.22 & -0.25\\
HD 44007    & 0.25 & 0.24 & 0.11 & 0.10 & -0.27 & -0.05 & -0.17 & -0.18\\
HD 63791    & 0.27 & 0.29 & 0.16 & 0.24 & -0.33 & -0.27 & -0.31 & -0.29\\
HD 74462    & \nodata & 0.09 & -0.03 & 0.01 & \nodata & -0.13 & -0.34 & -0.34\\
\enddata					     
\tablenotetext{*}{ The abundance for Ba {\small II} is taken from Aoki et al. (2002)}
\end{deluxetable}
\clearpage

\begin{deluxetable}{lrrrrr}
\tablecolumns{6}
\tablewidth{0pc}
\tablecaption{Sensitivity of Abundances to Atmospheric Parameters: HD
  94028 \label{error1}}
\tablehead{
\colhead{} & \colhead{$\Delta$T} & \colhead{$\Delta$log $g$}& \colhead{$\Delta\xi$}
& \colhead{$\Delta$[Fe/H]} & \colhead{}\\ \colhead{Element} &
\colhead{+125 K} & \colhead{+0.5 dex}
& \colhead{+0.5 km s$^{-1}$} & \colhead{+0.2 dex} & \colhead{Total}}
\startdata
[Fe/H] & 0.11 & 0.12 & -0.11 & 0.01 & 0.20 \\
\[[Mg/Fe] & -0.03 & 0.09 & 0.07 & 0.00 & 0.12 \\
\[[Ca/Fe] & 0.00 & 0.09 & 0.02 & 0.00 & 0.09 \\
\[[Cr/Fe] & 0.00 & -0.11 & 0.07 & -0.01 & 0.13\\ 
\[[T{\small II}/Fe] & -0.06 & -0.24 & -0.04 & 0.01 & 0.25\\
\[[Ba {\small II}/Fe] & -0.01 & -0.15 & -0.09 & 0.01 & 0.18
\enddata					     
\end{deluxetable}
\clearpage

\begin{deluxetable}{lrrrrr}
\tablecolumns{6}
\tablewidth{0pc}
\tablecaption{Sensitivity of Abundances to Atmospheric Parameters:
BD+37 1458 \label{error2}}
\tablehead{
\colhead{} & \colhead{$\Delta$T}   & \colhead{$\Delta$log $g$}   & \colhead{$\Delta\xi$}
& \colhead{$\Delta$[Fe/H]} & \colhead{}\\ \colhead{Element} &
\colhead{+125 K} & \colhead{+0.5 dex}
& \colhead{+0.5 km s$^{-1}$} & \colhead{+0.2 dex} & \colhead{Total}}
\startdata
[Fe/H] & 0.12 & -0.06 & -0.12 & -0.01 & 0.17 \\
\[[Mg/Fe] & -0.04 & -0.06 & -0.01 & 0.00 & 0.08 \\
\[[Ca/Fe] & -0.03 & -0.01 & -0.03 & 0.00 & 0.05 \\
\[[Cr/Fe] &  0.01 & 0.06 & 0.09 & 0.00 & 0.11\\
\[[T{\small II}/Fe] & -0.07 & 0.23 & 0.04 & 0.03 & 0.22\\
\[[Ba {\small II}/Fe] & -0.02 & 0.17 & -0.13 & 0.04 & 0.16
\enddata					     
\end{deluxetable}
\clearpage

\begin{deluxetable}{lrrrrr}
\tablecolumns{6}
\tablewidth{0pc}
\tablecaption{Sensitivity of Abundances to Atmospheric Parameters: CS
22892-052 \label{error3}}
\tablehead{
\colhead{} & \colhead{$\Delta$T}   & \colhead{$\Delta$log $g$}   & \colhead{$\Delta\xi$}
& \colhead{$\Delta$[Fe/H]} & \colhead{}\\ \colhead{Element} &
\colhead{+100 K} & \colhead{+0.3 dex}
& \colhead{+0.3 km s$^{-1}$} & \colhead{+0.2 dex} & \colhead{Total}}
\startdata
[Fe/H] & 0.14 & -0.06 & -0.19 & -0.01 & 0.22 \\
\[[Mg/Fe] & -0.03 & -0.04 & 0.05 & 0.01 & 0.08 \\
\[[Ca/Fe] & -0.05 & 0.05 & 0.15 & 0.01 & 0.15 \\
\[[T{\small II}/Fe] & -0.09 & 0.19 & -0.05 & 0.01 & 0.14\\
\[[Ba {\small II}/Fe] & -0.01 & 0.10 & -0.17 & -0.01 & 0.19
\enddata					     
\end{deluxetable}
\clearpage

\begin{deluxetable}{lrrrrrrr}
\tablecolumns{7}
\tabletypesize{\footnotesize}
\tablewidth{0pc}
\tablecaption{Program Star Abundances with Dereddened Atmospheres\label{dereddened}}
\tablehead{
\colhead{Star ID} & \colhead{[Fe/H]} & \colhead{[Mg/Fe]} &
\colhead{[Ca/Fe]} &\colhead{[Cr/Fe]} & \colhead{[TiII/Fe]} &
\colhead{[BaII/Fe]} & \colhead{[$\alpha$/Fe]\tablenotemark{\dag}}}
\startdata				
BS 16076-006 &	-3.00	& 0.43	&\nodata& \nodata & 0.28 & $<$-0.56 &  0.36   \\
BS 16077-007 &	-2.82	& 0.21	& 0.20  & \nodata & 0.34 & $<$-0.58 &  0.25   \\
BS 16080-054 &	-2.74	& 0.68	& 0.32  & \nodata & 0.87 & -0.19    &  0.62   \\
BS 16080-093 &	-2.73	& 0.35	& 0.20  & \nodata & 0.59 & 0.11	    &  0.38   \\
BS 16084-160 &	-2.93	& 0.34	& 0.11  & \nodata & 0.28 & $<$-1.71 &  0.24   \\
BS 16467-062 &	-3.79	& 0.50	&\nodata& \nodata & 0.49 & $<$ 0.09 &  0.50   \\
BS 16470-073\tablenotemark{*}&	-0.42	&\nodata& 0.47  &  0.07	  & 0.06 & -0.57    &  0.27   \\
BS 16470-092 &	-1.83	& 0.56	& 0.39  & \nodata & 0.41 & -0.42    &  0.45   \\
\enddata				     	      
\tablenotetext{*}{Eu probable, 4129.2 EuII line is present but could not be measured.}
\tablenotetext{\dag}{Average of Mg, Ca, and TiII}
\tablenotetext{\ddag}{Likely metal rich, all lines measured $>$ 200 m\AA}
\end{deluxetable}
\clearpage

\begin{deluxetable}{lrrrrrr}
\tablecolumns{7}
\tabletypesize{\footnotesize}
\tablewidth{0pc}
\tablecaption{Program Star Abundances with no reddening correction in Atmospheres\label{reddened}}
\tablehead{
\colhead{Star ID} & \colhead{[Fe/H]} & \colhead{[Mg/Fe]} &
\colhead{[Ca/Fe]} &\colhead{[Cr/Fe]} & \colhead{[TiII/Fe]} &
\colhead{[BaII/Fe]}}
\startdata		
BS 16076-006 &	-3.16	& 0.46	&\nodata& \nodata & 0.32 & $<$ -0.54 \\
BS 16077-007 &	-2.82	& 0.21	& 0.20  & \nodata & 0.34 & $<$ -0.58 \\
BS 16080-054 & 	-2.74	& 0.69	& 0.32  & \nodata & 0.87 & -0.18   \\
BS 16080-093 & 	-2.78	& 0.37	& 0.21  & \nodata & 0.57 & -0.14   \\
BS 16084-160 & 	-2.94	& 0.34	& 0.13  & \nodata & 0.27 & $<$ -1.72 \\
BS 16467-062 &	-3.79	& 0.50	&\nodata& \nodata & 0.49 & $<$ 0.09  \\
BS 16470-073 & 	-0.83	&\nodata& 0.67  & -0.09   & -0.10& -0.78   \\
BS 16470-092 & 	-1.95	& 0.61	& 0.42  & \nodata & 0.45 & -0.42   \\
\enddata				     	      
\end{deluxetable}

\end{document}